\documentclass[10 pt, a4paper]{article}
\usepackage{graphicx}
\usepackage{siunitx}

\usepackage{setspace}

\usepackage{lineno}

\usepackage{hyperref}
\hypersetup{hidelinks}

\usepackage[hang,flushmargin]{footmisc}

\usepackage[backend=bibtex8,style=nature,autocite = superscript]{biblatex}
	\DeclareBibliographyCategory{citedmain}
	\AtEveryCitekey {\ifnum\therefsegment=1
		{\addtocategory{citedmain} {\thefield{entrykey}}}\else\fi}
	
\bibliography{TULW-arxiv}

\usepackage[printwatermark]{xwatermark}
\usepackage{xcolor}
\usepackage{graphicx}

\newwatermark[allpages,color=gray!20,angle=90,scale=2,xpos=60,ypos=0]{DRAFT}

\begin{document}

\begin{spacing}{1} 
	\begin{center}
{\Large \textbf{{Transversal ultrasound light guiding \\ deep into scattering media}}}

\vspace{5pt}
{\large {Maxim Cherkashin,$^{1,*}$ Carsten Brenner,$^{1}$ Georg Schmitz,$^{2}$ \\ and Martin Hofmann$^{1}$}}
\end{center}  
\end{spacing}
\footnotetext[1]{Photonics and Terahertz Technology, Ruhr University Bochum, 44780, Bochum, Germany; 
$^2$Medical Engineering, Ruhr University Bochum, 44780, Bochum, Germany; $^*$Correspondence to: \href{mailto:maxim.cherkashin@rub.de}{maxim.cherkashin@rub.de}}
	\vspace{-10pt}
\begin{spacing}{1} 
\begin{quotation}
{
		Biomedical applications	requiring tissue diagnosis, activation, and treatment could be substantially leveraged by optical methods, owing to their unique feature set.
			However, their widespread application is severely limited by the strong light scattering that occurs in many tissues of interest, which dramatically limits achievable penetration depths.
				Here we demonstrate a new method to solve this issue by utilizing free-running ultrasound waves, transversal to the light propagation direction, to guide light into deeper tissue regions.
					We study the formation of the ultrasound-induced refractive index structures and waveguides using simple ultrasound field configurations and analyze their effects on the propagation of short light pulses. 
						Our results show waveguide support and associated light intensity increase up to the depths of at least~\SI{20}{\milli\meter} in Intralipid-20\% phantoms with a reduced scattering coefficient close to real tissue. 
							Thus we present an important milestone towards low-loss light delivery, focusing and manipulation in deep tissue.
	}
\end{quotation}
	\vspace{3pt} 
\newrefsegment

Optical methods are safe for biomedical applications and provide high speed, high resolution, and selectivity.\autocite{Baker2010} 
	By virtue of these advantages, various optical concepts are currently used for light-based tissue imaging,\autocite{Ntziachristos2010} activation~\autocite{Park2017} and treatment.\autocite{Sandell2011}

Their practical applicability, however, is severely impeded by light scattering.
	Light scattering hinders both the ability to focus light into tissue, and to deliver light into regions deeper than a few millimeters,\autocite{Ntziachristos2010,Jacques2013a} thus hindering the ability to image, activate and cure deep tissue.
	
Over time, several strategies were developed to mitigate or circumvent these difficulties.
	Passive strategies include utilization of near-infrared light as opposed to visible to access biomedical transparency window(s), or usage of sophisticated filtering concepts for imaging, as, for instance, in ballistic imaging, optical coherence tomography, and diffuse optical tomography.\autocite{Ntziachristos2010, Haisch2012c}

Several active strategies were also introduced to handle light scattering and its effects by influencing light propagation.
	Firstly, the amount of scattering events can be diminished, using invasive solutions\autocite{Simandoux2015a, Park2017} and optical clearing,\autocite{Tuchin2007a, Genina2010} or affecting the distribution of scatterers.\autocite{Chan1996}  

To address deep tissue light delivery \textit{in vivo}, here we propose a method to perform light waveguiding in scattering media using refractive index gradients induced by transversally traveling ultrasound waves. 
	We demonstrate waveguide formation and associated light intensity increase, tracing the path towards low loss light delivery, focusing and manipulation in deep tissue. 

\begin{figure}[tb] 
	\centering
	\includegraphics[width=89mm]{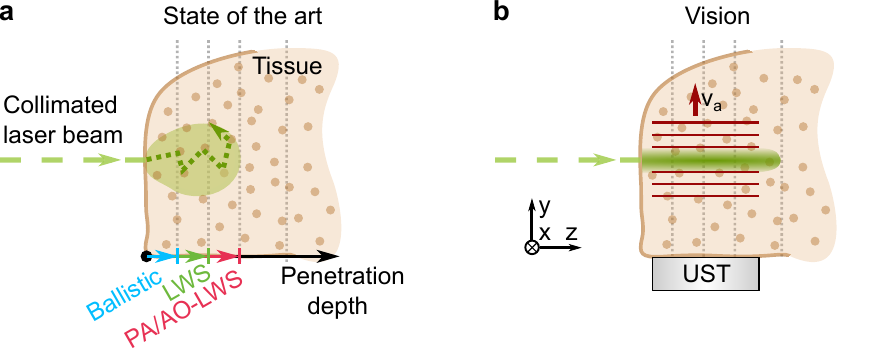}
	\caption{					
		{Light transport in highly scattering media and low loss light delivery mediated by ultrasound.}  \label{fig:VisionZero}
			\textbf{a}, Collimated laser light beam impinging on the highly scattering tissue forms a typical forward-scattering pattern: straight line propagation is broken, and light penetration depth is limited. 
				\textbf{b}, Light can be guided deeper into the tissue by carefully crafted ultrasound field, emitted by ultrasound transducer~(UST), and detrimental effects of scattering are reduced. 
		}
\end{figure}

Among the most successful recent active approaches used to counteract light scattering and improve light delivery is light wavefront shaping~(LWS).\autocite{Vellekoop2007,Mosk2012} 
	This technique requires the knowledge of light intensity at some point in depth as a feedback signal to optimize incident light field and improve light delivery to the given position.
	
Hence, it is possible to employ photoacoustic~(PA) and acoustooptic~(AO) interactions to recover light intensity information from the deep tissue.
	Subjected to lower propagation losses of ultrasound as compared to light, both of the modalities are able to increase the sensing range and thus are able to extend the depths where light wavefront shaping methods may be used.\autocite{Xu2011, Chaigne2013b, Lai2015a}
		However, even PA- and AO-aided light wavefront shaping~(PA/AO-LWS) require that threshold illumination levels are achieved first. 
			Therefore, their application towards deeper tissue regions is also dependent on achievements in light delivery.

Recently, the concept of tunable acoustic gradient~(TAG) lenses~\autocite{McLeod2006,McLeod2007,McLeod2007a} has been successfully applied to turbid media.\autocite{Chamanzar2019}
	This demonstrates, that standing ultrasound~(US) waves in a closed cavity provide sufficient AO-interaction and may aid light delivery in-depth.

In this paper we demonstrate that needle-like light propagation into deeper regions in scattering media may be achieved non-invasively using light waveguiding by transient transversal ultrasound fields~({Fig.~\ref{fig:VisionZero}}). 
	We present the idea and confirm its applicability towards biological tissues using proof-of-principle experiments with pulsed illumination and free-running ultrasound waves in a bulk scattering medium. 
		We study simple ultrasound field configurations and show maximum light intensity increase in depths up to 9 penetration depths in phantom media with estimated reduced scattering coefficient of $\mu_s'\simeq$~\SI{4.5}{\centi\meter^{-1}} for \SI{532}{\nano\meter} laser light.
			Fundamentally, this research is paving the way to low-loss light delivery, light focusing and highly customized light manipulation deep in scattering media.

Our concept builds up on the following blocks: A) an ultrasound wave can change the refractive index of the medium~\autocite{Pitts1999b, Pitts2000a}~({Fig.~\ref{fig:A}}), B) arbitrary ultrasound patterns can be created using wave interference~\autocite{Melde2016}~({Fig.~\ref{fig:B}}), C) transversal gradients of refractive index can guide the light,\autocite{Yang2012, Shi2015} D) ultrasound propagation is not altered substantially in scattering media compared to water~\autocite{Selfridge1985, Szabo2014}~({Fig.~\ref{fig:D}}). 
	Based on these premises {our concept is to use transient ultrasound waves} transversal to the light propagation direction {to perform light waveguiding} deeper in bulk scattering medium~({Fig.~\ref{fig:E}}).

Each of the statements above (A--D) raises no concern or doubt in its validity, when considered separately, however, to introduce our concept gradually and support the {premises} described, we present our results in the mentioned order.

\subsubsection*{Planar refractive index structures}
Ultrasound is propagating in the medium as a longitudinal wave. 
	That means, areas of local compression and rarefaction of the material follow each other and move together in the waves' propagation direction. 
		These  zones of local density change are altering the refractive index of the medium.\autocite{Pitts1999b} 

Within linear US emission and propagation, the pressure produced by the ultrasound transducer is proportional to the driving voltage.
	Then, in a linear approximation,\autocite{Pitts1999b} the local change of the refractive index $\Delta n$ is proportional to the local ultrasound pressure $p$, {$\Delta n \propto \gamma p$}, where the piezo-optic coefficient of water $\gamma=1.2 \ldots 1.5\times10^{-10}$~\si{\pascal^{-1}}.

This process forms a dynamic ultrasound-induced refractive index structure~(US-IRIS) in the medium. 
	In the case of a typical few-cycle sine wave excitation of a planar transducer, this creates sinusoidal refractive index gradients, traveling across the medium with US speed.

\begin{figure}[tb] 
	\centering 	
	\includegraphics[width=120mm]{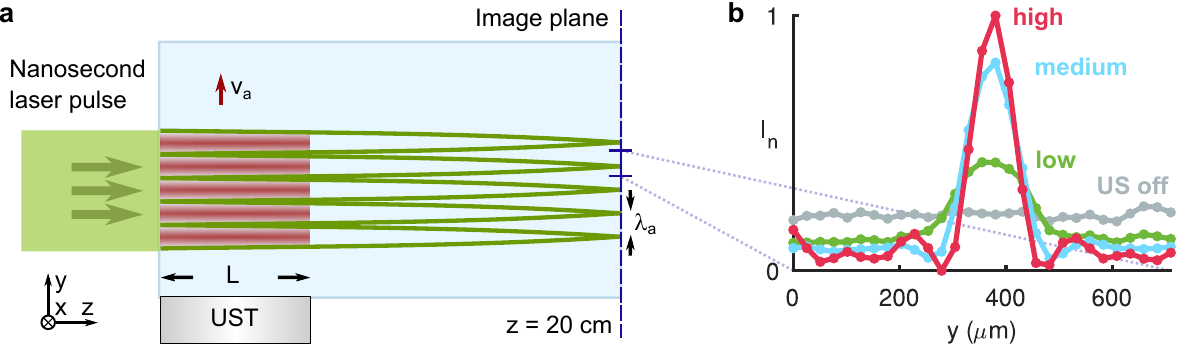}
		\caption{				
			{Ultrasound-induced lens array formation and associated pressure-dependent light intensity redistribution.}   \label{fig:A}
				\textbf{a}, Planar ultrasound transducer~(UST) is excited by a few-cycle sine wave, which induces a transversal refractive index profile, acting as cylindrical lens array for a wide collimated light beam input~(top view). 
					\textbf{b}, Change of the light intensity distribution in response to increase of the US pressure illustrated in the beam cross-section as captured by a camera.
						Lateral extent of the shown area corresponds to a single full cycle of sine wave used for UST excitation.
							Full dynamic illustration of the light focusing in \textbf{b} is presented in Supplementary Movie 1.
		}
\end{figure}

Since the speed of light is much higher than that of the US, the resulting movement of the US-IRIS while the short light pulse passes through it is negligible. 
	Hence, if a sufficiently short light pulse is sent sideways, it is impinging on a quasi-static transversal refractive index distribution~({Fig.~\ref{fig:A}a}). 
		Then, if US waves traveling along $y$ direction, at the moment of illumination $t=t_p$, $\Delta n(y)|_{t=t_p} \propto \textrm{sin}(2 \pi y/\lambda_a + \phi)$,  where $\lambda_a=v_a/f_a$ is acoustic wavelength, determined by US speed in the medium $v_a$ and its frequency $f_a$, and the phase $\phi$ is dependent on the synchronization between the light pulse and the ultrasound burst. 

For a collimated light beam, a single full cycle of such US-IRIS~(termed hereafter ''single channel'') is comparable in its effect to a cylindrical lens. 
	The optical pathlength through the structure is then dependent on its current location, amplitude of the refractive index change, and the extent of the US-IRIS in the direction of light propagation~(z-axis in~{Fig.~\ref{fig:A}a}). 
		Therefore, an alteration of the ultrasound pressure leads to the respective change of the refractive power of the US-IRIS. 
			Thus, such system is capable of one-dimensional light focusing, where the position of the focal plane is tunable via ultrasound pressure. 

{Fig.~\ref{fig:A}b} illustrates this effect for a single channel of the plane-wave~(1D) US-IRIS. 
	To capture the light intensity redistribution along the beam's cross-section in response to the US pressure increase, a camera is used in water-based experiments as a detector with a pulsed frequency-doubled Nd:YAG laser~(\SI{532}{\nano\meter}, \SI{5}{\nano\second} pulse width) used as a light source~(see Methods). 
		In {Fig.~\ref{fig:A}b}, the US amplitude is increased to align the focal plane of the 1D US-IRIS with the focal plane of the camera.

The following observations could be made: first, increasing the US pressure suppresses the background light intensity. 
	This is in accordance with light redistribution from the area of the clear aperture of the single channel of the US-IRIS to its focal spot. 
		Additionally, when the foci of US-IRIS and the camera objective align~(highest US pressure in {Fig.~\ref{fig:A}b}), the light intensity distribution shows a characteristic Airy pattern, corresponding to the typical point spread function of a lens~(see also {Fig.~\ref{fig:B}a}). 
			Full dynamic illustration of the outcomes of the US pressure tuning in {Fig.~\ref{fig:A}b} is presented in  the Supplementary Movie~1. 

Accordingly, when wide beam illumination is used, multiple focal lines are produced. 
	Then, the line spacing is determined by the acoustic wavelength of the applied ultrasound in the medium $\lambda_a$, and the US-IRIS operation is similar to a microlens array, as illustrated in {Fig.~\ref{fig:A}a}.

To summarize, a single channel of a transversal plane-wave~(1D) US-IRIS acts as a pressure-tunable cylindrical lens.

\subsubsection*{Wave interference for US-IRIS shaping}

Having one-dimensional light focusing by the planar US-IRIS confirmed, it is possible to use the interference of two or more ultrasound waves to create more complex pressure profiles and corresponding refractive index distributions.
	Two simple US field configurations of practical interest are crossed and column-focused arrangements. 

\begin{figure} [tb] 
	\centering  
	\includegraphics[width=120mm]{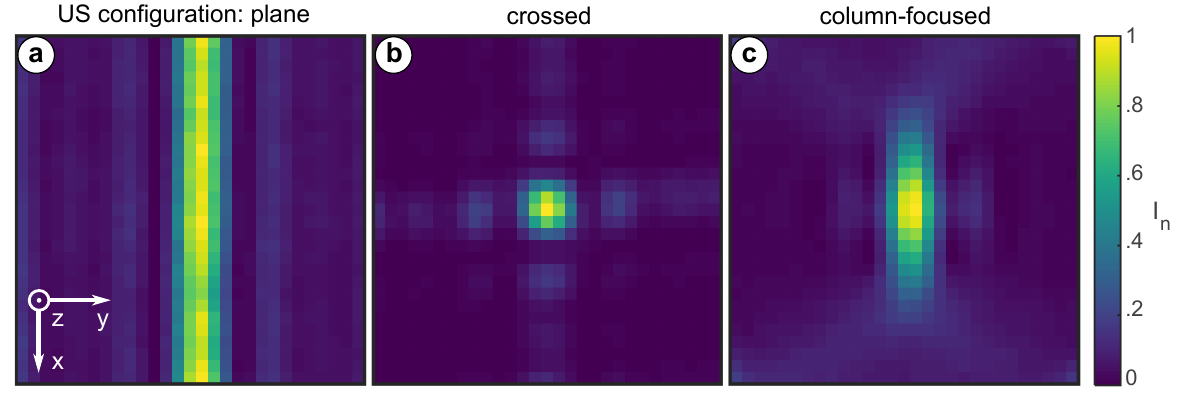} %
	\caption{					
		{Single channel light intensity profile dependency on ultrasound field shape.} \label{fig:B}
			En face 2D snapshots of light intensity distributions produced in their focal plane by planar(\textbf{a}), crossed(\textbf{b}) and column-focused (\textbf{c}) arrangements of the US fields in water as captured by the camera.
				Images FOV corresponds to 710~x~710~\si{\micro\meter}~($\lambda_a \times \lambda_a$). All panels share the colormap, with individual panels normalized before visualization. Focusing dynamics of \textbf{b} and \textbf{c}, similar to {Fig.~\ref{fig:A}b}, is illustrated in full in Supplementary Movies 2 and 3 respectively.
	}
\end{figure}

When a perpendicular combination of two planar transducers is used, 2D focusing of the light beam can be achieved within each channel of such 2D US-IRIS. 
	To illustrate this, {Fig.~\ref{fig:B}b} shows a focal plane image of a single channel of such a crossed US configuration. Again, the produced light intensity pattern corresponds to the point-spread function of a lens. 
		The shape and position of side lobes suggest a square aperture function, which is in agreement with the US configuration used. 
			Similarly to the planar US-IRIS, when wide beam illumination is used, such a crossed US arrangement creates a 2D grid of focal points, spaced by the acoustic wavelength $\lambda_a$ in both $x$ and $y$ directions.

To better utilize the available US pressure, an acoustic lens might be used to achieve a column-focused ultrasound configuration. 
	Since the refractive power of the US-IRIS is proportional the US pressure, higher refractive index contrasts can be achieved in this way.
		The light intensity distribution produced in such case is illustrated by~{Fig.~\ref{fig:B}c}. 
			This pattern corresponds to an elliptic aperture function and is in accordance with the US configuration used. 

So far, we have demonstrated, that ultrasound waves traveling  in water transversally to the light beam propagation direction could manipulate its behavior and that their effect depends on the applied US pressure and the US configuration. 

Accordingly, changing the applied ultrasound pressure will affect the refractive power of the US-IRIS and correspondingly shift its focal plane. 
	That means we could expect the light intensity distributions shown in {Fig.~\ref{fig:B}a-c} {to be translated} along the light propagation axis~(z in {Fig.~\ref{fig:A}a}). 
		Since the strength of the light-ultrasound interaction depends on the US pressure, a high-pressure column-focused US as used to produce {Fig.~\ref{fig:B}c} is preferred. 

\subsubsection*{US-IRIS in scattering phantoms}
Since the acoustic properties of soft tissues do not differ significantly from those of water,\autocite{Szabo2014} we could expect comparable effects. 
	Therefore, to further study the effects of US-IRIS on light propagation in scattering media we explored the light intensity distribution in scattering phantoms. 

Intralipid-20\% dilutions are common liquid scattering phantoms for light transport studies. 
	They are proven to be long-time stable and to provide inter-comparable results.\autocite{DiNinni2011,Spinelli2014}

However, reconstruction of the light distribution in the scattering phantoms with complex geometry presents a substantial challenge. 
	In homogeneous and axially layered samples, simulation methods,\autocite{Wang1995} or recently developed non-invasive techniques of light fluence assessment are available.\autocite{Hussain2014,Schreurs2017} 
		However, transversal structuring of the medium, induced by the ultrasound, extends beyond their current applicability.

Therefore, to perform high-resolution 3D mapping of light intensity in the depth of a bulk scattering sample, we utilized raster scanning of the volume-of-interest by a fiber-based point detector.\autocite{Cherkashin2018} 
	In this way it is possible to locally assess the available light intensity (see Methods). 

To illustrate the effects of the ultrasound field presence on the light propagation in the scattering medium, we further analyze the US on/off maximum light intensity ratios in the beam cross-section. 
	Since US fields are reproduced well in low concentration water-based Intralipid-20\% dilutions,\autocite{Cook2011} and we expect the focal plane intensity distribution to be preserved, these results shall resemble {Fig.~\ref{fig:B}c}.
 
 \begin{figure} [tb] 
 	\centering       
 	\includegraphics[width=120mm]{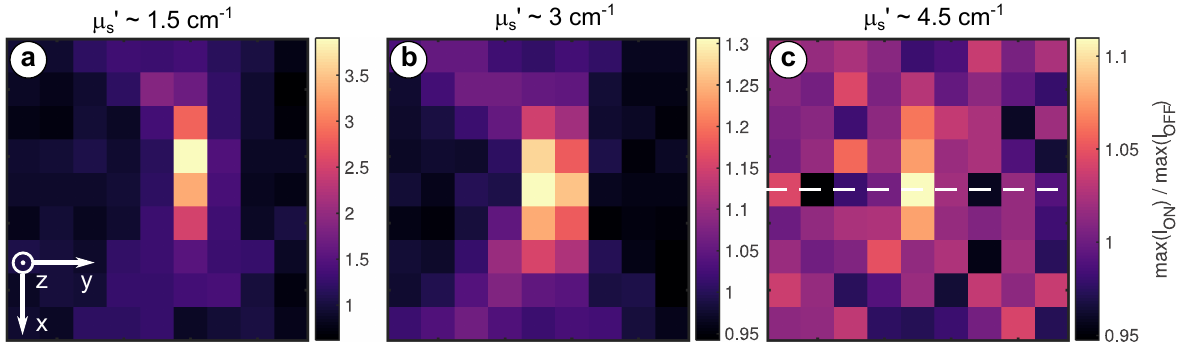}
 		\caption{				
 			{Variation of US-induced light intensity redistribution in scattering phantoms.} \label{fig:D}
	 			US on/off ratio of maximum light intensity distributions over the beam cross-section sampled with fiber probe~(see Methods) in scattering phantoms with reduced scattering coefficient $\mu_s'$ estimated to be \textbf{a},~\SI{1.5}{\centi\meter^{-1}}, \textbf{b},~\SI{3.0}{\centi\meter^{-1}}, \textbf{c},~\SI{4.5}{\centi\meter^{-1}} respectively for a \SI{532}{\nano\meter} light incidence. Imaging depth of 2~cm corresponds respectively to 3~(\textbf{a}), 6~(\textbf{b}), and 9~(\textbf{c}) penetration lengths.  
		 			Images FOV corresponds to 400x400~\si{\micro\meter}, scan step is \SI{50}{\micro\meter}, en face view. All panels share the colormap, however, the scale is adjusted for each panel. 
				 		Dashed line in \textbf{c} shows the cross-section direction for {Fig.~\ref{fig:E}a}.
		}
 \end{figure}
 
Such US on/off maximum light intensity ratio captures are presented in {Fig.~\ref{fig:D}a-c} for phantom media with estimated reduced scattering coefficients $\mu_s'$ of \SI{1.5}{\centi\meter^{-1}}, \SI{3}{\centi\meter^{-1}}, and \SI{4.5}{\centi\meter^{-1}} respectively  for a laser wavelength $\lambda=$~\SI{532}{\nano\meter}. 
	The imaging depth $z=$~\SI{2}{\centi\meter} corresponds then to 3, 6, and 9 penetration depths respectively.
		The expected similarity of the patterns in {Fig.~\ref{fig:D}} to the central portion of the {Fig.~\ref{fig:B}c} is notable, with gradual degradation of the light confinement degree with increased scattering. 
			We see this similarity as a proof that US pressure distribution and its effect on the light propagation are preserved up to the added amount of scatterers.

The raise of the light scattering leaves only a small core region of the increased maximum light intensity in the most extreme scattering case studied~({Fig.~\ref{fig:D}c}). 
	This result confirms that the area of a higher light intensity is restricted in the beam cross-sectional plane. 
		Thus, the US-IRIS ability to locally increase the available light intensity in scattering medium is proved. 

\subsubsection*{Deep light waveguiding by transversal US}
Finally, to prove the waveguiding capabilities of the US-IRIS, transversal restriction of the light propagation has to be confirmed. 
	Therefore, it is necessary to evaluate the light intensity distribution inside the US-IRIS as well. 

\begin{figure}[tb] 
	\centering
	\includegraphics[width=89mm]{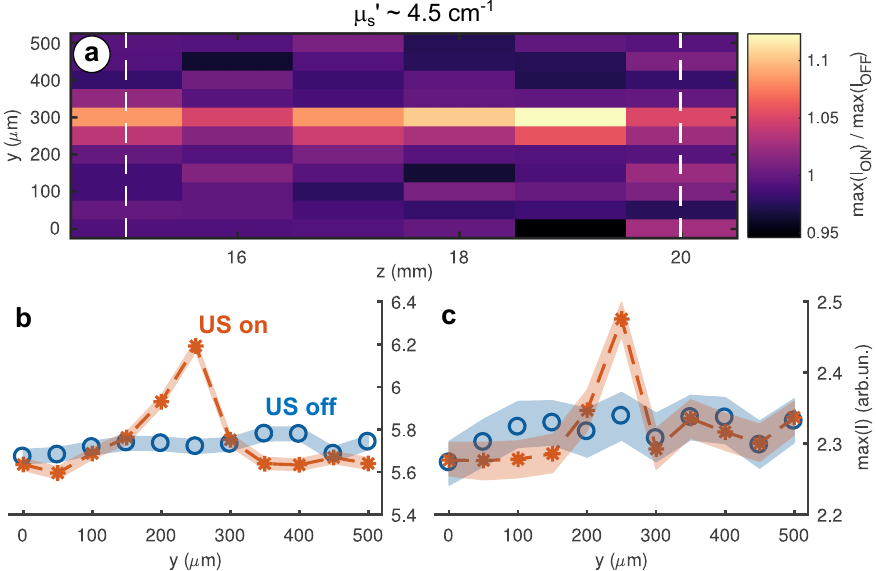} 
	\caption{							  
		{Cross-section of light intensity distribution within ultrasound-induced refractive index structure and single line US on/off comparisons.} \label{fig:E}
			\textbf{a}, Maximum light intensity is increased along the core of the US-induced waveguide in scattering phantom with $\mu_s'\sim$~\SI{4.5}{\centi\meter^{-1}}, as captured with a fiber probe, top view~(see Methods). 
				Depth of $z=$~\SI{20}{\milli\meter} corresponds to 9 penetration depths for laser wavelength of \SI{532}{\nano\meter}. 
					Dashed lines show the locations represented in the panels \textbf{b} and \textbf{c}, respectively. 
						\textbf{b-c} Maximum light intensity without~(open circles) and with US-induced light waveguiding~(asterisks) demonstrating increase of maximum light intensity at respective locations over the diffused light background. 
								Shaded areas correspond to $\pm1.96$ times standard error of the mean for $N_{acq}=100$ pulses used for averaging.
		}
\end{figure}

{Fig.~\ref{fig:E}a} shows the US on/off maximum light intensity ratio over the direction of the light propagation and column-focused US-IRIS extent. 
	A higher intensity area is notable in the central part of the figure, corresponding to the waveguide axis. 
		Therefore, the waveguiding capabilities of the US-IRIS in the scattering medium are confirmed.

The corresponding US on/off maximum light intensity comparison in the beginning and at the end of the scanned region, indicated by the dashed lines in {Fig.~\ref{fig:E}a} is presented in {Fig.~\ref{fig:E}b-c} along with statistical information. 
	This data further demonstrates the relation of the light intensity on the waveguide axis to the diffused light background. 
		Quite notably, the distributions, corresponding to the US-IRIS effects are similar to the light intensity in the focal plane of the lens {Fig.~\ref{fig:A}b}.

As following from {Figs.~\ref{fig:D}-\ref{fig:E}}, a column-focused US-IRIS provides 2D localized point light delivery and supports light waveguiding up to at least 9 penetration depths in scattering phantoms with estimated $\mu_s'\simeq$~\SI{4.5}{\centi\meter^{-1}} for a laser wavelength of \SI{532}{\nano\meter}. 

\subsubsection*{Towards TULW \textit{in vivo}}
The maximum level of $\mu_s'$, tested experimentally is approaching the lower boundary of real tissue parameters.\autocite{Jacques2013a} 
	Practical~(i.e. \textit{in vivo}) application of the transversal ultrasound light waveguiding~(TULW) may require higher US pressures to deal with even higher scattering. 
		This can be readily achieved in practice, using margins built in by robust experimental design~(see Methods).

For future applications of our approach, there are several possible strategies to optimize the US pressure delivery. 
	First, the duty cycle and waveform of the transducer excitation could be optimized. 
		To avoid overheating of the transducer and ensure stable operation conditions, the peak excitation voltage applied in our experiments was kept below 30~\% of the rated maximum. 
			Second, ultrasound coupling between the US delivery and the experimental volume can be improved greatly. This is easy to achieve since acoustic impedances of soft tissues and water are matching well. 
				Moreover, when considering possible optogenetic applications of TULW, the skull possesses better US transmission properties~\autocite{Szabo2014} than the glass wall used in our experiments.
					Finally, better acoustic lenses could be used to provide tighter focusing into a high-pressure column(s).

These three means of further US field optimization provide a bridge towards application of TULW in real tissues, to partially offset the light losses due to scattering and to access deeper tissue regions.
	Given the dominance of forward scattering in the tissues~\autocite{Jacques2013a} and the low NA of the induced waveguides, we envisage the creation of non-invasive, ultrasound-induced optical needles for deep tissue activation and sensing.\autocite{Simandoux2015a,Rogers2013,Yuan2015}

Our investigation shows, that TULW acts on both light focusing in turbid media~({Fig.~\ref{fig:D}c}) and light delivery into deep tissue regions~({Fig.~\ref{fig:E}}). 
	Relying on free-running ultrasound and equipped with advantages of the ultrasound field engineering, highly complex temporally and spatially restricted US-IRIS can be created, limited in their existence, extent and configuration only by ultrasound propagation properties.
		Therefore, tailored light intensity arrangements in depth can be set up on demand to help light-driven biomedical methods against scattering.

Hence, by investigating the basic US-IRIS configurations, we open the possibility to use the advantages of the ultrasound engineering to aid light propagation in scattering media, based on TULW, or higher complexity concepts.\label{endofmain}
\end{spacing}
\vspace{-15pt}
\printbibliography[title={References},category=citedmain]
{\small
\noindent\textbf{{Supplementary Information}}
Methods, Safety concerns and Related work sections as well as Extended data Figures 1--3 and Supplementary Movies 1--3 are presented online.
\vspace{3pt}

\noindent{\textbf{Acknowledgements}}
The authors acknowledge the help of W.~D. Putro through early stage of the project, support of T.~Kranemann and T.~Ersepke with hydrophone measurements, discussions with H.~Estrada, and manuscript proofreading by C.~Saraceno.  This project was partly funded by EU FP7 People: Marie Curie Actions Programme, Award No. 317526 (OILTEBIA).
\vspace{3pt}

\noindent
{\textbf{Author Contributions}}
M.~C. and M.~H. conceived the idea. M.~C. and C.~B. designed the experiments. M.~C. performed the experimental work. M.~C. and C.~B. analyzed the data. G.~S. supervised the ultrasound part of the work. M.~H. performed general supervision of the project. M.~C., C.~B., and M.~H. wrote the manuscript. All authors discussed the results and revised the manuscript. 
\vspace{3pt}

\noindent
\textbf{Author Information}
The authors declare no competing interests.
Correspondence and requests for materials should be addressed to  M.~C.~(\href{mailto:maxim.cherkashin@rub.de}{maxim.cherkashin@rub.de}).
\vspace{-2pt}

\noindent
\textbf{Data Availability}
Data is available from the corresponding author upon reasonable request. 
\vspace{3pt}

\noindent
\textbf{ORCiD} M.~C.:~\href{https://orcid.org/0000-0003-1283-7476}{0000-0003-1283-7476},
 C.~B.:~\href{https://orcid.org/0000-0002-8764-9261}{0000-0002-8764-9261},
  G.~S.:~\href{https://orcid.org/0000-0001-5876-7202}{0000-0001-5876-7202}.
  }
\newpage
\newrefsegment
\begin{spacing}{1}
\section*{\centering Methods}
\subsection*{Experimental arrangement}

A transmitter-channel-detector scheme is typically used to investigate the effects of the medium parameters on light transport. 
	We describe the components of our experiments following this order.

The light source is a frequency-doubled pulsed Nd:YAG laser~(Minilite II, Coherent) providing \SI{5}{\nano\second} light pulses with a repetition rate of \SI{14}{\hertz} at wavelength of \SI{532}{\nano\meter}. 
	The light beam is either expanded with a 5x beam expander for experiments requiring wide field-of-view~({Figs.~\ref{fig:A}-\ref{fig:B}}) or focused with a low NA lens~($f=$~\SI{30}{\centi\meter}) for scattering medium experiments~({Figs.~\ref{fig:D}-\ref{fig:E}}). 

A tri-section glass-made container~(float glass, \SI{3}{\milli\meter} thickness), separated into frontal, main and a sideways auxiliary~(Aux) chambers is used to achieve the required US field configuration and to reduce the volume of the phantom media necessary for experiments~({Extended Data Figure~\ref{fig:ExtM}a}, top view). Light is propagating through the frontal and main chambers, while the sideways chamber is used to adjust US delivery.

A planar single element active ultrasound transducer~(UST, model A395S-SU, nominal frequency \SI{2.25}{\mega\hertz}, element diameter \SI{38}{\milli\meter}, Olympus Panametrics-NDT) is placed facing downwards in the auxiliary chamber~(emitting in positive \textit{x} direction). 
	The UST is excited at its specified spectral maximum with a 7-cycle \SI{2.09}{\mega\hertz} sine burst using a combination of a waveform generator and a linear RF amplifier, while the delay generator ensures synchronization between the system components. 
	
Plane US waves, emitted by the UST are redirected using a planar metallic reflector~(R1, refer to {Extended Data Figure~\ref{fig:ExtM}b-d} for an en face view). 
	The UST is offset in $-z$ direction against the boundary between the frontal and the main chambers.
		Additionally, the frontal chamber is kept empty.
			This configuration ensures minimal distance between the ultrasound-induced refractive index structure (US-IRIS) in the main chamber and the boundary of the phantom media, minimizes edge effects on the light entrance, and eliminates beam pre-shaping by the US wave. 

The auxiliary chamber is filled with purified water.
	The main chamber is filled either with purified water~({Figs.~\ref{fig:A}-\ref{fig:B}}) or volume-guided Intralipid-20\%~(IL) dilution~(see {Scattering phantoms} and {Extended Data Table~\ref{table:samples}}). 
	
Additionally a planar or parabolic US reflector is utilized to produce crossed or focused US fields as required~(R2, {Extended Data Figure~\ref{fig:ExtM}c-d} respectively). 
	In this way, the US-IRIS of desired configuration is created in the medium in the main chamber.

The light intensity distribution is picked up after passage through water with a camera placed inline behind the main chamber. 
	A set of neutral density~(ND) filters is used to avoid camera saturation. 
		{Fig.~\ref{fig:A}b and Fig.~\ref{fig:B}} were produced with camera acquisitions data and parabolic reflector configuration~({Extended Data Figure~\ref{fig:ExtM}d}), using necessary synchronization and wide illumination.

In scattering phantom experiments, a fiber probe is used for high resolution 3D raster scanning of the light intensity distribution. 
	The probe is a cleaved GIF625 fiber, the distal end of which is affixed in a metallic needle, connected to a linear stage assembly~(LSA).\autocite{Cherkashin2018}

Given the applied US frequency of \SI{2.09}{\mega\hertz} and the speed of sound of 1482~m/s,\autocite{Szabo2014} the acoustic wavelength is estimated to be \SI{710}{\micro\meter}. 
	Therefore, for the \SI{62.5}{\micro\meter} fiber core, its detector area is well below the periodicity of US-IRIS, and fiber is used as point detector. 
		Although this particular arrangement requires time-consuming raster scanning, it is possible to obtain 3D light intensity distribution within the US-IRIS.

The light picked up by the fiber probe in the medium is guided towards a fast photodetector, while another fast photodetector is used to monitor pulse-to-pulse variations. 
	Both signals are fed to a digital sampling oscilloscope~(DSO, 5~GHz sampling frequency). 
		The data acquired with fiber probe is used to produce {Figs.~\ref{fig:D}-\ref{fig:E}}.

System operation control, data acquisition, processing and analysis are performed using a PC running MATLAB~(Mathworks). 

\subsection*{Scattering phantoms}

To produce phantom media for experiments, different dilutions of Intralipid-20~\%~(IL, Sigma-Aldrich) are prepared.
	A calculated volume of IL is mixed with purified water with the aid of a magnetic stirrer to obtain a dilution of required volume and estimated reduced scattering coefficient $\mu_s'$. 
		The parameters of the IL dilutions used in the experiments are listed in  {Extended Data Table~\ref{table:samples}}.

To estimate the reduced scattering coefficients of the IL dilutions, we use available literature data on IL optical properties~\autocite{Michels2008,DiNinni2011,DiNinni2012,Spinelli2014} and assume linear dependence of the reduced scattering coefficient $\mu_s'$ of a dilution on IL volume fraction~(VF).

Provided in the {Extended Data Table~\ref{table:samples}} are estimated reduced scattering coefficients $\mu_s'$ of Intralipid-20~\% dilutions, computed as following: $\mu_s' = VF \cdot \mu_{sIL}'$. 
	The used value $\mu_{sIL}' \simeq 300$~\si{\centi\meter^{-1}} is reduced scattering coefficient attributed to Intralipid-20\%, measured at laser wavelength of \SI{543}{\nano\meter},\autocite{Michels2008} which is sufficiently close to the wavelength of \SI{532}{\nano\meter} used in our experiments.
	
The estimates above might underestimate the real values of the reduced scattering coefficient of the IL dilutions due to the influence of dependent scattering.\autocite{Michels2008, DiNinni2011, Raju2017}
	An alternative estimation is possible, based on the spectral properties of light extinction in IL dilutions and available near-infrared measurements.~\autocite{DiNinni2011,DiNinni2012}
		This provides that real values of the reduced scattering coefficient of the IL dilutions used are within a factor of 1 to 1.5 to the values presented above. 
			However, we adhere to conservative, lower-boundary estimates as presented in the {Extended Data Table~\ref{table:samples}}.
	
\subsection*{Data acquisition and processing}

The raw signals are averaged to decrease the effects of fluctuations and noise for both camera and fiber-based experiments.
	
To produce {Fig.~\ref{fig:A}b} and {Fig.~\ref{fig:B}}, $N_{acq}=30$ images are recorded for each US peak pressure level $p$, proportional to the input voltage. 
	Single images are then averaged, which forms an image stack used for visualizations.
		The data from the image stack is normalized for a given figure.
			The extended data is presented in {Extended Data Figures~\ref{fig:Ext1}}-{\ref{fig:Ext2}} and Supplementary Movies 1-3.								
		
To produce the US on/off comparisons~({Figs.~\ref{fig:D}-\ref{fig:E}}), the following procedure is used. 
	Two raster scanning acquisitions with a fiber-based probe are performed sequentially. 
		First, the light intensity is acquired with ultrasound burst synchronized to the respective position in the volume of interest~("US on"). 
			Second, the waveform generator driving US transducer is disabled and acquisition process is repeated~("US off"). 
				
For {Fig.~\ref{fig:D}},  $N_{acq}=98$ acquisitions are acquired for each raster scanning point, that are ensemble averaged at DSO and stored. 
	Then, maximum values of the averaged traces are extracted.
		Finally, point-by-point ratios of respective US on/off acquisitions are computed and plotted.
			
To provide statistical information presented in {Fig.~\ref{fig:E}b-c}, single acquisitions are obtained.
	Then, $N_{acq}=100$ sequential acquisitions are ensemble averaged, respective maximum values are extracted and presented along with maximum values of single traces in {Fig.~\ref{fig:E}b-c}, while their point-by-point ratios are used for {Fig.~\ref{fig:E}a}.

The data of the reference photodetector was used to monitor the pulse-to-pulse variations. 
	No anomalies were found between respective US on/off data pairs. 

\subsection*{Hydrophone measurements}

For a number of tests, hydrophone measurements were performed.
	In this way, the linearity of the transducer output against the input waveform is assured. 	
		Additionally, measurements of the peak US pressure in the main and in the auxiliary chambers are used to estimate the glass wall reflectivity. 

These measurements are performed using an end-cable calibrated system of 1~mm needle PVDF hydrophone with preamplifier and booster~(Precision Acoustics). 
	The hydrophone is placed in the respective part of the experimental container on the height of the laser beam. 
		To ensure perpendicular incidence of the ultrasound on the hydrophone membrane, a similar plane metallic reflector is utilized to fold the ultrasound beam. 
			In this way the ultrasound pressure in plane is captured. 
				The boosted hydrophone output is recorded using the same digital sampling oscilloscope~(DSO) as described above. $N_{acq}=100$ acquisitions are averaged.
				
The ratio of the peak pressure in the main and auxiliary chambers obtained this way is considered representative of the glass wall transmission. 
	The immersed glass wall transmission coefficient is then estimated to 20~\%, which is in the range reported in the literature.\autocite{Viennet2009}

It is important to note, that soft tissue and skull boundaries shall present better acoustic transmission properties,\autocite{Selfridge1985, Szabo2014} therefore, improving the ultrasound pressure delivery.

\section*{\centering Safety concerns}
When considering possible applications, it is important to conform to relevant safety limitations. 
	One of the key points of TULW is the pressure dependence of the effects' strength. 
		In conjunction with high scattering of tissues, that indicates that higher US pressures can be of interest. 
		
According to the hydrophone measurements performed~(see above), the maximum peak pressure in plane is measured to be 200~kPa~(auxiliary chamber) and 33~kPa~(main chamber). 
	Based on the light intensity redistribution in water with different configurations of US fields, we estimate a maximum pressure multiplication factor of $\beta=4$ in column-focused configuration with respect to a planar configuration.
		
		Therefore, the maximum peak pressure used in our experiments is estimated to be 800~kPa, when neglecting the glass wall transmission losses of 80~\%.
			
This pressure level is near to the lower boundary of the range used for diagnostic ultrasound.\autocite{Szabo2014}
	In total, improving the US delivery and increasing the applied pressure within safe levels, can offer an additional pressure and US-IRIS refractive power gain up to 30 times, that transversal ultrasound light guiding could benefit from.
	
Two other key ultrasound dosimetry characteristics, total emitted power and averaged power,\autocite{Szabo2014} can be adjusted by optimizing the duty cycle of the ultrasound excitation used. 
	Therefore, TULW is having potential for safe usage with even higher pressure levels.
	
\section*{\centering Related work}
The effect of refractive index alteration by a propagating acoustic wave is known for a long time in gases, liquids and solid media and has found numerous applications. In the final part of the paper we provide some of the most important connections to highlight the intersection of the fields where US-IRIS and TULW reside.

In this paper we exploited US-IRIS formed in liquid media. From this standpoint, our US-IRIS shows similarities to liquid core-liquid cladding (so-called L2) elements. Such L2 systems are known in the transformation optics domain, and have proved to possess interesting light manipulation capabilities.\autocite{Wolfe2004,Yang2012,Shi2015,Liu2017}

The analysis applied for L2 systems could be extended to US-IRIS as well. In particular, in comparison to L2 systems, whose optical properties are tuned by adjustment of interdiffusion of the used liquids, our US-IRIS is controlled with the US pressure distribution. Therefore, the refractive properties of the US-IRIS are dependent on frequency, amplitude, and speed of the employed ultrasound in the medium, as well as on its propagation conditions. Combination of these parameters will determine the transversal profile of the refractive index, which could be used to analyze and describe its' optical properties. 

Our experimental and {simulation results} agree well with such a description. In particular, we observed a notable similarity of our simulated and experimentally obtained intensity distributions to those of beam self-interference, {particular to multimode waveguides.\autocite{Yang2012}} Please refer to the extended data presented in {Extended Data Figures~\ref{fig:Ext1}}-{\ref{fig:Ext2}} and Supplementary Movies 1-3.

The similarity of the obtained light intensity distributions in the non-scattering and the scattering media tested~({Fig.~\ref{fig:D}}) confirms, that added amounts of scattering agent do not alter the mechanism of waveguide formation. However, the complex nature of light propagation in scattering media motivates further investigation of the impact of scattering on the waveguiding capabilities of TULW in order to fully exploit the potential of our approach. 

In particular, the question whether the scatterers are displaced along with a US wavefront, and how the scattering cross-section might alter in compressed and rarefied areas of the medium has to be addressed. 

The changes of the optical properties, induced by macroscopic tissue compression~\autocite{Chan1996,Tuchin2007a} suggest, that the ordering of the scatterers may have an impact on the total transmission of the tissues. 
	Therefore, ultrasound-coordinated changes in the scatterer density may add additional nonlinear contributions to further improve US-mediated light guiding, for instance, via light self-action in dense particle suspensions.\autocite{Bezryadina2017}

Finally, our concept is relying on the short pulse illumination of free-running ultrasound waves, resulting in momentary captures of the US pressure distribution in media. This makes it different from tunable acoustic gradient~(TAG) lenses and guides,\autocite{McLeod2006,McLeod2007,McLeod2007a} successfully applied to turbid media recently.\autocite{Chamanzar2019}

The usage of free-running ultrasound allows us to engineer the ultrasound pressure distribution to create highly temporally and spatially restricted waveguiding structures, principally limited in their existence, extent and configuration only by US propagation losses.  Moreover, the versatility of the ultrasound field engineering translates into a variety of more complex US-IRIS, that can be created. 
In particular, we would like to address the possibility of crossed-field (2D) US-IRIS to form axially extended waveguide grids. 

TULW is particularly well fit with photoacoustic and acousto-optic imaging towards tomographic scales. Both ultrasound and pulsed light subsystems are present in the mentioned cases, and operate in lower US frequency range. 

Moreover, advanced US field shaping with multielement transducers may be utilized to perform in-depth light wavefront shaping.

It is important to note, that our results counter the viewpoint of the negligibility of the refraction index modulation by US waves,\autocite{Wang2001a} kept over last decades.\autocite{Wang2004b,Elson2011,Resink2012c,Gunther2017} This particular type of light-ultrasound interaction is regaining interest~\autocite{Kobayashi2006, Resink2014, Jarrett2014, Chamanzar2019} and our approach presents a different viewpoint to enrich the understanding of the complex processes involved.

To conclude, by presenting and examining the US-IRIS and TULW, we show a proof of the concept and its applicability towards real tissues. 
	We showed isolated light guiding observation with little to no beam pre-shaping. 
		We have successfully demonstrated its' applicability up to 9 penetration depths in scattering medium with an estimated reduced scattering coefficient of $\mu_s'\simeq$~\SI{4.5}{\centi\meter^{-1}} and an incident laser wavelength of \SI{532}{\nano\meter}. 

We have presented the effects of the US-IRIS on the light propagation in scattering medium using light source in the visible range. 
	As shorter wavelengths are undergoing stronger scattering, compared to the near-infrared range, the latter region is often preferred for higher penetration and imaging depths.\autocite{Ntziachristos2010,Jacques2013a}
		However, TULW is not particularly dependent on the used laser wavelength.
			Therefore, its' combination with suitable near-infrared sources could further extend the penetration and sensing depths achievable.
				
In this way we enable further optimization of US light guidance and reassure that accessing the deeper tissue regions is possible.

\newpage
\section*{\centering Extended Data}
\setcounter{figure}{0} 
\renewcommand{\figurename}{Extended Data Figure}
\renewcommand{\tablename}{Extended Data Table}
\begin{figure}[htb]
	\setlength{\leftskip}{-1.5cm}
	\includegraphics[width=155mm]{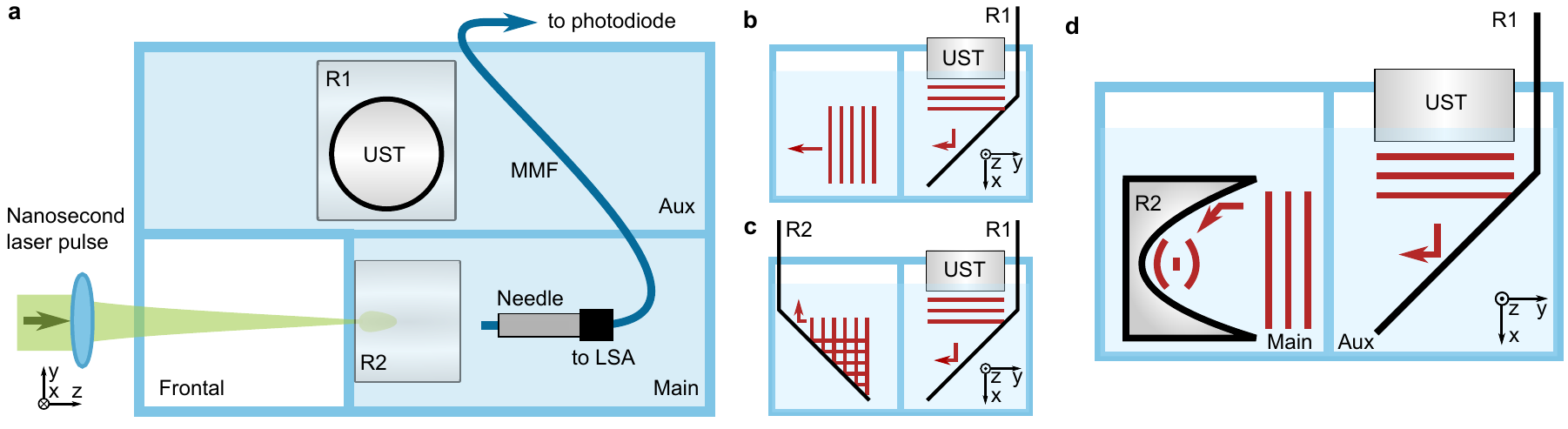} \\
	\caption{ 					
		{Experimental layout and configurations of ultrasound-induced refractive index structures.} \label{fig:ExtM} 
		\textbf{a}, top view of the experimental chamber, \textbf{b-d} en face views of the US reflector arrangements for planar~(1D, \textbf{b}), crossed~(2D, \textbf{c}) and column-focused~(\textbf{d}) US arrangements. UST - ultrasound transducer; R1, R2 - ultrasound reflectors; MMF - multimode fiber; LSA - linear stage assembly for 3D raster scanning. 
		Configurations \textbf{b-c} were used in preliminary work and provided to gradually introduce the concept. 
		Parabolic reflector arrangement in \textbf{d} is used throughout experiments described.  Respective synchronization adjustment and image windowing is applied to demonstrate the effects of US arrangements in \textbf{b} and \textbf{c}.  
		}
\end{figure}

\begin{table} [ht]
	\begin{center}
		\caption{{Dilutions used to produce {Figs.~\ref{fig:D}-\ref{fig:E}}.} \label{table:samples}}
		{\footnotesize
			\begin{tabular}{{c}|{c}|{c}|{c}}
				\hline
				\multicolumn{2}{c|}{Volume,~mL} &  &\\
				\cline{1-2} 
				Intralipid-20~\% 	&
				Purified water 	 	& 
				Volume fraction (VF) &
				Estimated $\mu_s'$, \SI{}{\centi\meter^{-1}} \\
				\hline
				5	&  	995	& 	0.005	& 1.5	\\ 
				10	&  	990	& 	0.01	& 3		\\ 
				15	&  	985	& 	0.015	& 4.5	\\ 
				\hline 
			\end{tabular} \\
			Total dilution volume: \SI{1}{\liter}.		
		}
	\end{center}
\end{table} 

\newpage
\begin{figure} [htbp] 
	\centering
	\includegraphics[width=90mm]{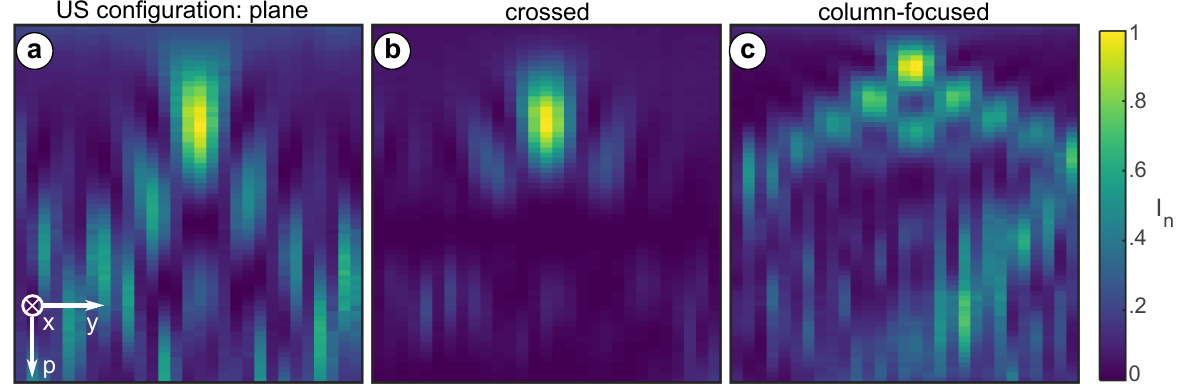} %
	\caption{						
		{Pressure-dependent light intensity redistribution dynamics, illustrated across single US-IRIS channel.} \label{fig:Ext1}
		Effects of planar(\textbf{a}), crossed(\textbf{b}) and column-focused (\textbf{c}) arrangements of the US fields in water as captured by the camera. 
			Images illustrate extended version of  {Fig.~\ref{fig:A}b} and are demonstrating light intensity along the middle of frames shown in {Fig.~\ref{fig:B}a-c}.
				Images extent over $y$ axis corresponds to \SI{710}{\micro\meter}~($\lambda_a$). All panels share the colormap, with individual panels normalized before visualization. 
					Video illustration of shown dynamics is presented in full in Supplementary Movies 1-3 respectively.
		}
\end{figure}
\newpage
\begin{figure} [htbp] 
	\centering
	\includegraphics[width=90mm]{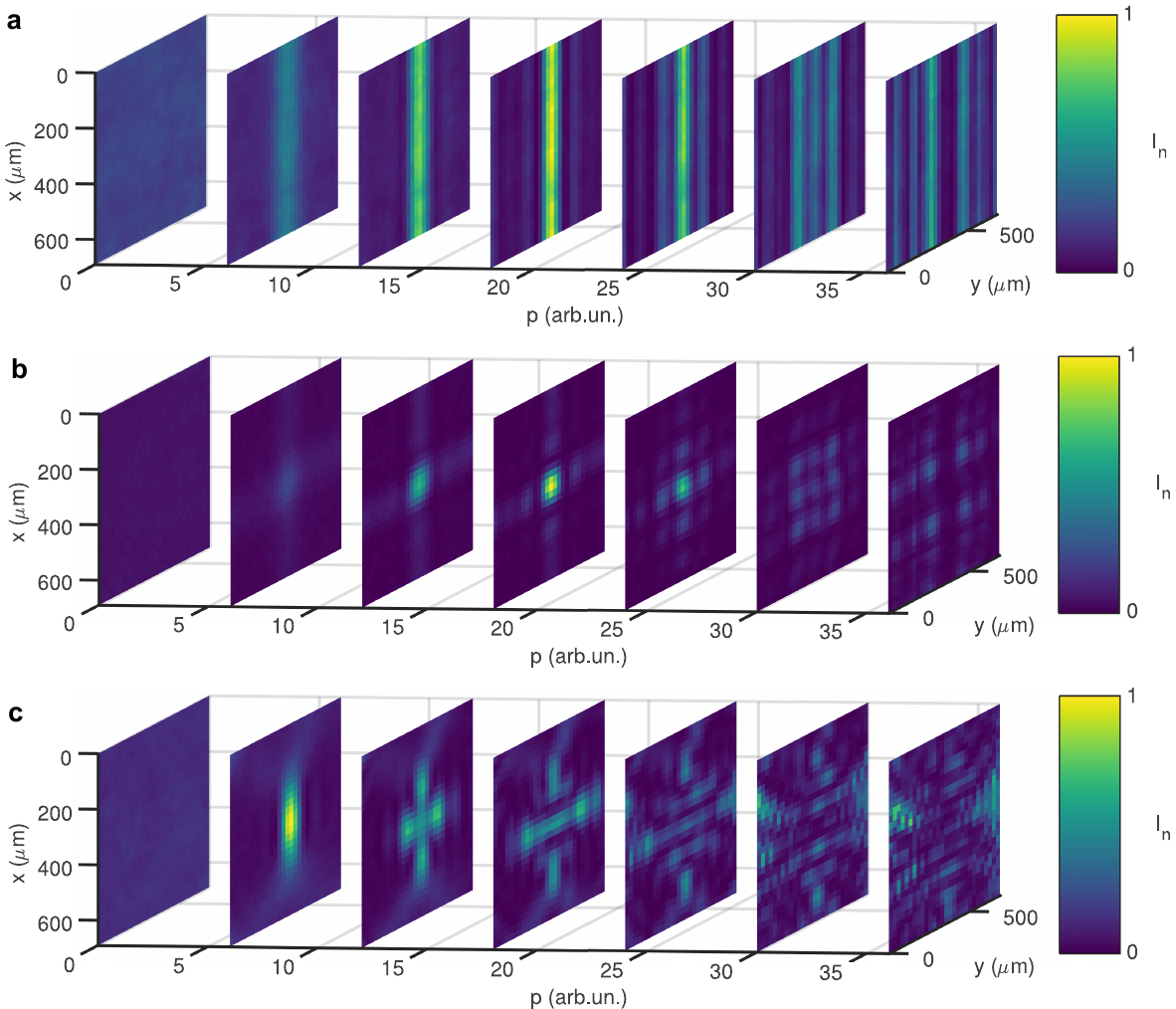} %
	\caption{					
		{Supplementary Movies 1-3 stills.} \label{fig:Ext2}
		Pressure-dependent light redistribution for planar(\textbf{a}), crossed(\textbf{b}) and column-focused (\textbf{c}) arrangements of the US fields in water respectively, as captured by the camera.
			Images field-of-view corresponds to 710$\times$710~\si{\micro\meter}~($\lambda_a \times \lambda_a$). 
				Individual panels are normalized before visualization.
	}
\end{figure}
\end{spacing}

\newpage
\newpage
\printbibliography[title={References (continued)},notcategory=citedmain]
\end{document}